 \documentstyle[prl,aps,epsf,amsfonts,amssymb,multicol,epsfig]{revtex}

\begin{document}
 \draft
 \title{Field Dependence of the  Josephson Plasma Resonance in Layered
 Superconductors with Alternating Junctions}
 \author{L.N. Bulaevskii$^1$, Ch. Helm$^2$
 }
 \address{$^1$Los Alamos National Laboratory, Los Alamos, NM 87545, USA \\
 $^2$Institut f{\"u}r Theoretische Physik,  ETH H{\"o}nggerberg, 
8093 Z{\"u}rich, Switzerland }
 \date{\today}
 \maketitle

\begin{abstract}
The  Josephson plasma resonance in  layered superconductors 
with alternating critical current densities is investigated in a low 
perpendicular magnetic field. In the vortex solid phase the current
densities and the squared bare plasma frequencies decrease linearly with the
magnetic field. Taking into account the coupling due to charge fluctuations on the
layers, we extract from recent optical data for SmLa$_{1-x}$Sr$_x$CuO$_{4-\delta}$  
the Josephson penetration length $\lambda_{ab} \approx 1100$ \AA \ \ 
parallel to the layers at $T=10$ K. 
\end{abstract}



\begin{multicols}{2}

The suppression of the Josephson interlayer coupling and consequently
the Josephson plasma resonance (JPR) frequency, $\omega_0$,
due to misaligned pancake vortices in layered 
superconductors with identical intrinsic junctions was calculated 
in the Refs.~\onlinecite{Kosh,Kosh0,kb,lnb} in  good agreement with
experimental data in the Bi$_2$Sr$_2$CaCu$_2$O$_{8-\delta}$ superconductor. 
It was found that in the vortex crystal phase, at low fields,  $\omega_0$ 
drops linearly with the vortex concentration (with the magnetic field
$B$ 
applied along the $c$-axis), 
while in the vortex liquid and glass state, at high fields, $\omega_0$ drops as 
$1/\sqrt{B}$ because vortices are uncorrelated along the $c$-axis and 
many of them contribute to the suppression of the Josephson coupling at a 
given point.  

Recently the JPR was studied by optical means
\cite{marel1,pimenov} 
in the layered 
superconductor SmLa$_{1-x}$Sr$_x$CuO$_{4-\delta}$ with
alternating intrinsic 
Josephson junctions containing SmO and LaO as  insulating barriers
between 
superconducting CuO$_2$-layers.  Reflectivity and transmission
measurements were performed with the incidence of light parallel to the layers
($ab$-plane) and the polarization of the electric field in $c$-direction. 
Then the effective dielectric function $\epsilon_{{\rm eff}}(\omega)$ 
was extracted from the data by using the Fresnel formulae. 
In crystals with $x=0.2$ two peaks in the loss function,  
$L(\omega)={\rm Im}[-1/\epsilon_{{\rm eff}}(\omega)]$, were found at
 frequencies 
6.6 and 8.9 cm$^{-1}$ at $B=0$.  These frequencies drop with $B$
linearly at 
low $B<0.5$ T and as $1/\sqrt{B}$ at $B>1$ T. 
The behavior of the JPR frequencies at high magnetic fields was attributed
to the vortex liquid state. 

The relative intensities 
of these peaks and the dependence of the JPR frequencies on $B$ 
at high fields were explained in the Refs.~\onlinecite{marel2,helm,ourprb} in
the model of two alternating intrinsic junctions with different Josephson
critical current densities. Thereby it is essential to take into account the charge 
coupling of neighboring junctions, i.e. the $c$-axis spatial dispersion of
Josephson plasmons \cite{art,tach,wir}, in order to explain the peak
amplitudes both for parallel and grazing incidence on the surface 
 \cite{marel2,helm,ourprb,ourepl}. 
The dimensionless parameter $\alpha$ characterizing the change of the chemical 
potential of superconducting layers with the electron concentration was
estimated from the field dependence of the JPR resonances in the loss
functions \cite{helm}.  

In this paper we will calculate the behavior 
of the JPR frequencies at low magnetic fields. 
This allows us to extract another important parameter of the 
SmLa$_{1-x}$Sr$_x$CuO$_{4-\delta}$ superconductor, 
the London penetration length $\lambda_{ab}$.  

We consider a crystal with alternating Josephson critical current densities
$J_m$, $m=1,2$, and corresponding bare plasma frequencies 
$\omega_{0m}^2=8\pi^2csJ_m/\epsilon_0\Phi_0$, where $s$ is the 
interlayer spacing 
which we assume to be similar in both junctions (for
SmLa$_{1-x}$Sr$_x$CuO$_{4-\delta}$  the difference 
is about 1.5\%) and $\epsilon_0$ is the high 
frequency dielectric constant ($\approx 19$ for 
SmLa$_{1-x}$Sr$_x$CuO$_{4-\delta}$, 
see Ref.~\onlinecite{helm}). We calculate first the dependence of the
bare JPR frequencies $\omega_{0m} (B)$ on $B$ without charge coupling
between the junctions.  Then we will account for the JPR dispersion due to 
$\alpha$ to find the
real JPR eigenfrequencies $\omega_m$ and the field dependence of 
the resonances $\omega_m$ in the loss function $L(\omega)$. 

Let us consider the field dependence of the JPR frequencies in 
the single vortex regime at low fields 
$B\ll B_{Jm}, B_{\lambda}$, where 
$ B_{J} =  \Phi_0/\lambda_{Jm}^2$, $B_{\lambda} = \Phi_0/4 \pi \lambda_{ab}^2$ and
$\lambda_{Jm}=
(\Phi_0cs/8\pi^2\lambda_{ab}^2J_m)^{1/2}$ are the Josephson lengths
of the  junctions of type $m=1,2$. In this limit the interaction between the
vortices is screened (single vortex regime). 
We neglect pinning assuming that mainly thermal fluctuations are 
responsible  for the meandering of the vortex line. 
This assumption is valid at sufficiently high temperatures $T$.  
In Ref.~\onlinecite{kb} it was shown that 
in the single vortex regime the  field dependence is given as 
\begin{eqnarray}
&&\omega_{0m}^2(B)=\omega_{0m}^2(0)(1-B/B_{0m}), \ \ B_{0m}=\Phi_0/I_m, 
\label{omegac} \\
&&I_m=\int d{\bf r}\langle[1-\cos\varphi_{n,1;n+1-m,2}({\bf r})]\rangle, 
\end{eqnarray}
where ${\bf r}$ is the in-plane coordinate, 
$\varphi_{n,1;n-m+1,2}({\bf r})$ is the phase difference 
between the layers $n1$ and $n-m+1,2$ and $\langle \dots \rangle$ is the
average over thermal disorder. 
We introduce the displacements $u_{nm}$ of the vortex from the 
straight line along the $c$-axis in the layers $nm$ and its Fourier transforms 
$u_m(q)$ with respect to the 
coordinate $n$ of the unit cell consisting of two different junctions, 
$0\leq q\leq 2\pi$. Then we 
express the phase difference via vortex displacements in quadratic 
approximation as it was done in Ref.~\onlinecite{kb} and obtain 
\begin{eqnarray}
I_1&=&\frac{\pi}{2}\left\langle\int \frac{dq}{2\pi}(1-\cos q)
|u_1(q)-u_2(q)|^2\times  \right.
\nonumber \\
&&\left.\ln\left(\frac{3.72\lambda_{J1}^2}{(u_{n1}-u_{n2})^2(1-\cos
q)}\right)\right\rangle
, \\
I_2&=&\frac{\pi}{2}\left\langle\int \frac{dq}{2\pi}(1-\cos q)
|u_1(q)-u_2(q)e^{iq}|^2\times \right.
\nonumber \\
&&\left.\ln\left(\frac{3.72\lambda_{J2}^2}{(u_{n1}-u_{n-1,2})^2(1-\cos
q)}\right)\right\rangle. 
\end{eqnarray}
In the following we will use the self-consistent harmonic approximation
(SCHA) replacing $(u_{n1}-u_{n2})^2$ by its average value 
$\langle(u_{n1}-u_{n2})^2\rangle=4r_{w1}^2$, where $r_{wm}$ is the
meandering length for junctions of the type $m$.  $I_m$ in 
terms of $r_{wm}$ is 
\begin{equation}
I_m=(\pi/2)r_{wm}^2\ln(0.8\lambda_{Jm}/r_{wm}).
\end{equation}
The increase of the vortex energy due to displacements is 
\begin{eqnarray}
&&{\cal E}_{\rm vor}=\frac{1}{2}\sum_q[E_{J1}|u_1(q)-u_2(q)|^2 + \\
&&E_{J2}|u_1(q)-u_2(q)e^{iq}|^2+W_M(|u_1(q)|^2+|u_2(q)|^2)],
\nonumber
\end{eqnarray}
where $E_{Jm}=\Phi_0J_m/2c$ is the Josephson 
coupling density in a junction of the type $m$ multiplied by the 
factor $(\pi/2)$
and $W_M=\Phi_0^2s/(4\pi \lambda_{ab}^2)^2$ accounts for the cage 
potential due to the 
nonlocal magnetic interaction between pancakes in different 
layers. After diagonalization of this energy we obtain the free energy 
functional of the vortex line 
\begin{eqnarray}
{\cal
F}&=&\frac{1}{2}\sum_{q,i=1,2}[E_i(q)|v_i(q)|^2-2T\ln|v_i(q)|^2], \\
E_{1,2}(q)&=&E_{J1}+E_{J2}+W_M\pm \\
&&(E_{J1}^2+E_{J2}^2+2E_{J1}E_{J2}\cos q)^{1/2}. \nonumber
\end{eqnarray}
After minimization with respect to the new variables $v_i(q)$ we obtain
the 
free energy as 
\begin{equation}
F(E_{Jm},W_M,T)=2T\sum_{q}\ln[E_1(q)E_2(q)]+F_0(T), 
\end{equation}
For $I_m$ we derive (neglecting logarithmic factors) 
\begin{eqnarray}
&&I_1=\frac{\pi r_{w1}^2}{2}=\frac{\pi}{2}
\frac{\partial F}{\partial E_{J1}}\approx
\frac{\pi T}{E_{J1}}f(E_{J1}), 
\nonumber \\
&&f(E_{J1}) =\left[1-\frac{1+2{\cal E}_{J2}}
{\sqrt{(1+4\overline{{\cal E}}+
4{\cal E}_{J1}{\cal E}_{J2})(1+4\overline{{\cal E}})}}\right],
\label{caseI} 
\end{eqnarray}
and similar for $I_2$, where ${\cal E}_{Jm}=E_{Jm}/W_M=
(\lambda_{ab}/\lambda_{Jm})^2=(\lambda_{ab}^2/\lambda_{cm}s)^2$
and 
$\overline{{\cal E}}=
({\cal E}_{J1}+{\cal E}_{J2})/2$. The values $\lambda_{cm}$ are 
determined by the zero field bare 
plasma frequencies $\omega_{0m} (B=0) =c/\sqrt{\epsilon_0}\lambda_{cm}$ in the
absence of an external magnetic field $B$.

For high fields $B\gg B_{Jm}, B_{\lambda}$ the vortex lattice is more dense, 
the interaction between the vortices diminishes pancake fluctuations, while
the enhanced tilt stiffness of the lattice favours larger fluctuations, i.e. a
reduced Josephson energy. 
We write down the energy functional in the Fourier representation ${\bf k}$ 
with respect to the in-plane coordinate ${\bf r}$ and the unit cell index
$n$. We separate transverse, 
$u_{t}({\bf k},n)$, and longitudinal, $u_{l}({\bf k},n)$, vortex lattice
displacements (for details see Ref. \onlinecite{kb}) 
\begin{eqnarray}
{\cal E}_{\rm vor}&=&{\cal E}_t(u_t)+{\cal E}_l(u_l), \\
{\cal E}_t(u_t)&=&\sum_{{\bf k},n,m}[C_{66}k^2+\Phi_{44}({\bf
k},m)]
|u_{t}({\bf k},2n+m)|^2, \\
{\cal E}_l(u_l)&=&\sum_{{\bf k},n,m}[\Phi_{11}+\Phi_{44}({\bf
k},m)] |u_{l}({\bf k},2n+m)|^2, 
\end{eqnarray}
where the summation over ${\bf k}$ is limited to the first Brillouin zone, 
which we approximate by the circle $k<K_0$, $K_0^2=4\pi B/\Phi_0$.
Further, 
$\Phi_{11}\approx (B^2s/4\pi\lambda_{ab}^2)(1-k^2/4K_0^2)$ is the 
compression stiffness, 
$C_{66}=A_{66}B\Phi_0s/(8\pi\lambda_{ab})^2$ is the shear modulus.
The parameter $A_{66}<1$ describes the fluctuation suppression of $C_{66}$,
cf. Ref. \onlinecite{c66}, 
which we approximate as $A_{66}=1-0.4B/B_{\rm melt}$ with $B_{\rm melt}$ being the
melting field, and 
\begin{equation}
\Phi_{44}(k,m)= E_{cm}+ E_{Jm}\eta_m
\end{equation}
is the tilt stiffness.  The cage energy is given as 
\begin{equation}
E_{cm}=
\frac{B\Phi_0s}{2(4\pi\lambda_{ab}^2)^2}\ln\left(0.5+\frac{0.13a^2}{r_{wm}^2}
\right) .
\label{cage}
\end{equation}
Here $a^2=\Phi_0/B$ is the intervortex distance and 
\begin{equation}
\eta_m=\frac{B}{2\Phi_0}\ln\frac{0.11 a^2}
{r_{wm}^2(1-0.53k^2/K_0^2)^2}+ \frac{4\pi }{a^4k^2}. 
\label{def}
\end{equation}

After diagonalization of the energy functional we obtain the eigenvalues
$E_{t1}$ 
and $E_{t2}$ for the transverse displacements with 
\begin{eqnarray} 
E_{t1}E_{t2}&=&S_{t1}S_{t2}+(S_{t1}+S_{t2})(E_{J1}\eta_1+E_{J2}\eta_2)+
\nonumber \\
&&2E_{J1}E_{J2}\eta_1\eta_2(1-\cos q), \\
S_{tm}&=&C_{66}k^2+E_{cm}.
\end{eqnarray}
For longitudinal eigenvalues $E_{lm}$ we need to replace  $S_{tm}$ with
$S_{lm}=\Phi_{11}+E_{cm}$ . 

Finally we find the free energy and by differentiating it with respect 
to $E_{Jm}$ we find the self-consistency equations for $r_{wm}$ as 
\begin{eqnarray}
&&I_m=\pi r_{wm}^2/2=\pi(r_{wt,m}^2+r_{wl,m}^2)/2, 
\label{caseII}
\\ 
&&r_{wt,1}^2= \frac{\Phi_0}{B} T \int\frac{d{\bf k}dq}
{(2\pi)^3}~\frac{[2E_{J2}\eta_2(1-\cos q)+
S_{t1}+S_{t2}]\eta_1}{E_{t1}E_{t2}} \nonumber \\
&&= \frac{\Phi_0}{B} \int\frac{d{\bf k}}
{(2\pi)^2}~\frac{T}{E_{J1}}\left[1-\frac{S_{t1}S_{t2}+(S_{t1}+S_{t2})
E_{J2}\eta_2}{\sqrt{D_t^2-(2E_{J1}E_{J2}\eta_1\eta_2)^2}}\right],
\nonumber \\
&&D_t=S_{t1}S_{t2}+(S_{t1}+S_{t2})(E_{J1}\eta_1+E_{J2}\eta_2)+
\nonumber \\
&&2E_{J1}E_{J2}\eta_1\eta_2, \label{last} \nonumber
\end{eqnarray}
and similar for $r_{wt,2}^2$ and $r_{wl,m}^2$. The 
Eqs.~(\ref{def})-(\ref{last}) should be solved self-consistently to find
the meandering lengths $r_{wm}$.

The bare frequencies $\omega_{0m}(B)$ determine the resonance frequencies 
$\omega_m(B)$, which are renormalized due to the charge coupling of the layers, 
and are observed experimentally in reflectivity, transmissivity and in the loss 
function $L (\omega)$. Next we derive  the renormalized frequencies and compare them 
with the measurement of these quantities in 
Ref.~\onlinecite{pimenov}. 
The parameter $\alpha=(\epsilon_0/4\pi e s)(\partial \mu/\partial {\rho})$ 
characterizes the interlayer coupling due to charge fluctuations \cite{art,tach,wir},
where  $\mu$ and $\rho$ are the chemical 
potential and charge density on the layers, respectively.  
The parameter $\alpha$ was estimated as 0.4 for
SmLa$_{1-x}$Sr$_x$CuO$_{4-\delta}$ at $x=0.2$ in Ref.~\onlinecite{helm} both
from the magnetic field dependence of the JPR resonances in the vortex liquid
state and from the relative amplitude of the resonances in $L(\omega)$ for $B=0$.  
For incident light perpendicular to the crystal surface $ac$ the wave
vector, $k_x$, of the wave propagating into the crystal determines the 
reflection coefficient, 
$R(\omega)$, according to the Fresnel formula, 
$R=|(1-n)/(1+n)|^2$, 
with the refraction index  $n(\omega)=ck_x(\omega)/\omega$.  In
Ref.~\onlinecite{helm,ourprb}
the 
dispersion relation $k_x(\omega)$ was calculated using the Maxwell
equations 
and the equation for the phase differences in a stack of intrinsic
Josephson 
junctions coupled  inductively and due to charge variations inside the layers.
The result is 
\begin{eqnarray}
&&\frac{c^2k_x^2}{\omega^2\epsilon_{0}}=
\frac{\epsilon_{\rm eff}(w)}{\epsilon_{0}}=
\frac{r(w-v_1)(w-v_2)+iS}{r w^2-(1+r)(2\alpha+1/2)w+iS_1}, 
\label{r}\\
&&v_m=\frac{\omega^2_m}{\omega^2_{01}}, \ \ 
v_{1,2}=(1+r)(1+2\alpha)\frac{1\mp \sqrt{1-p}}{2r},  
\label{v} \\
&&r(B)=\frac{\omega_{01}^2}{\omega_{02}^2}<1, \ \ 
p=\frac{4r(1+4\alpha)}{(1+r)^2(1+2\alpha)^2},  
\label{p} \\
&&S_1=w^{3/2} r (2 \alpha + 1/2) ({\tilde \sigma}_1 + 
{\tilde \sigma}_2) , \\
&&S=w^{1/2} [(2\alpha+1) r w (\tilde{\sigma}_1+\tilde{\sigma}_2)-
(1+4\alpha)(\tilde{\sigma}_1+\tilde{\sigma}_2 r)], \nonumber
\end{eqnarray}
where $\omega_m(B,\alpha) $
are the normalized JPR frequencies, $w=\omega^2/\omega_{01}^2$, 
$\tilde{\sigma}_m=4\pi\sigma_m/\epsilon_0\omega_{01}$, and $\sigma_m$
are the $c$-axis quasiparticle conductivities of the junctions.  
We see that the loss function $L (\omega)$ has two peaks at $\omega=\omega_m$ 
corresponding to zeros of $\epsilon_{{\rm eff}}(\omega)$ in the 
absence of dissipation.  These resonances in the loss function
correspond to the transverse plasma modes propagating along the layers.  
There is another characteristic frequency defined by the relation 
\begin{equation}
\omega_{\rm pole}^2(B)=[\omega_{01}^2(B)+\omega_{02}^2(B)](2\alpha+1/2), 
\label{omegat}
\end{equation}
which corresponds to a pole of ${\rm Im}(\epsilon_{{\rm eff}}(\omega))$ in the 
absence of dissipation, cf. Eq.(\ref{r}).  This frequency is near the minimum
of the loss function $L(\omega)$. 
This frequency is often called "transverse plasmon" 
and it is well defined experimentally as the peak in the real part of the 
optical conductivity, ${\rm Im} (\omega \epsilon_{\rm eff})$, 
see Ref.~\onlinecite{pimenov}. 

Next we calculate the field dependence of the 
frequency $\omega_{\rm pole}$, expressed by Eq.~(\ref{omegat}) 
via the field 
dependence of the bare frequencies $\omega_{0m}$,
Eq.~(\ref{omegac}), 
which we already found.  We obtain 
\begin{equation}
\omega_{\rm pole}^2(B)=\omega_{\rm pole}^2(0)(1-B/B_p) . 
\end{equation}
In the single vortex regime  ($B \ll B_{Jm}, B_{\lambda}$) the result is 
\begin{eqnarray}
B_p&=&\frac{\Phi_0^3 \epsilon_0  (\omega_{01}^2 (0) +
\omega_{02}^2 (0)  )}
{32\pi^3 c^2 s T  g({\cal E}_{J1},{\cal E}_{J2})}, \\
g({\cal E}_{J1},{\cal E}_{J2})&=&(1/2)[f(E_{J1})+f(E_{J2})] \\
&=& 1-\frac{1+
\overline{{\cal E}}}{\sqrt{(1+4\overline{{\cal E}}+
4{\cal E}_{J1}{\cal E}_{J2})(1+4\overline{{\cal E}})}}.
\end{eqnarray}

\begin{figure}
\begin{center} 
\epsfig{file=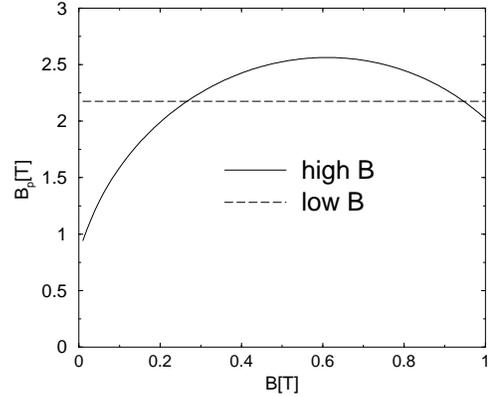,width=0.3\textwidth,clip=,angle=-90}
\end{center}
\caption{Field scale  $B_{p} (B)$ in the low ($B \ll B_{Jm}, B_{\lambda}$) or
 high  ($B \gg B_{Jm}, B_{\lambda}$) field limit  using
  Eq.~(\ref{caseI}) (dashed) or Eq.~(\ref{caseII}) 
(solid, $B_{\rm melt} = 2$ T) for $\lambda_{ab}=1700$ \AA, $T=2.3$ K. 
\label{Bcomp}
}
\end{figure}

\begin{figure}
\begin{center} 
\epsfig{file=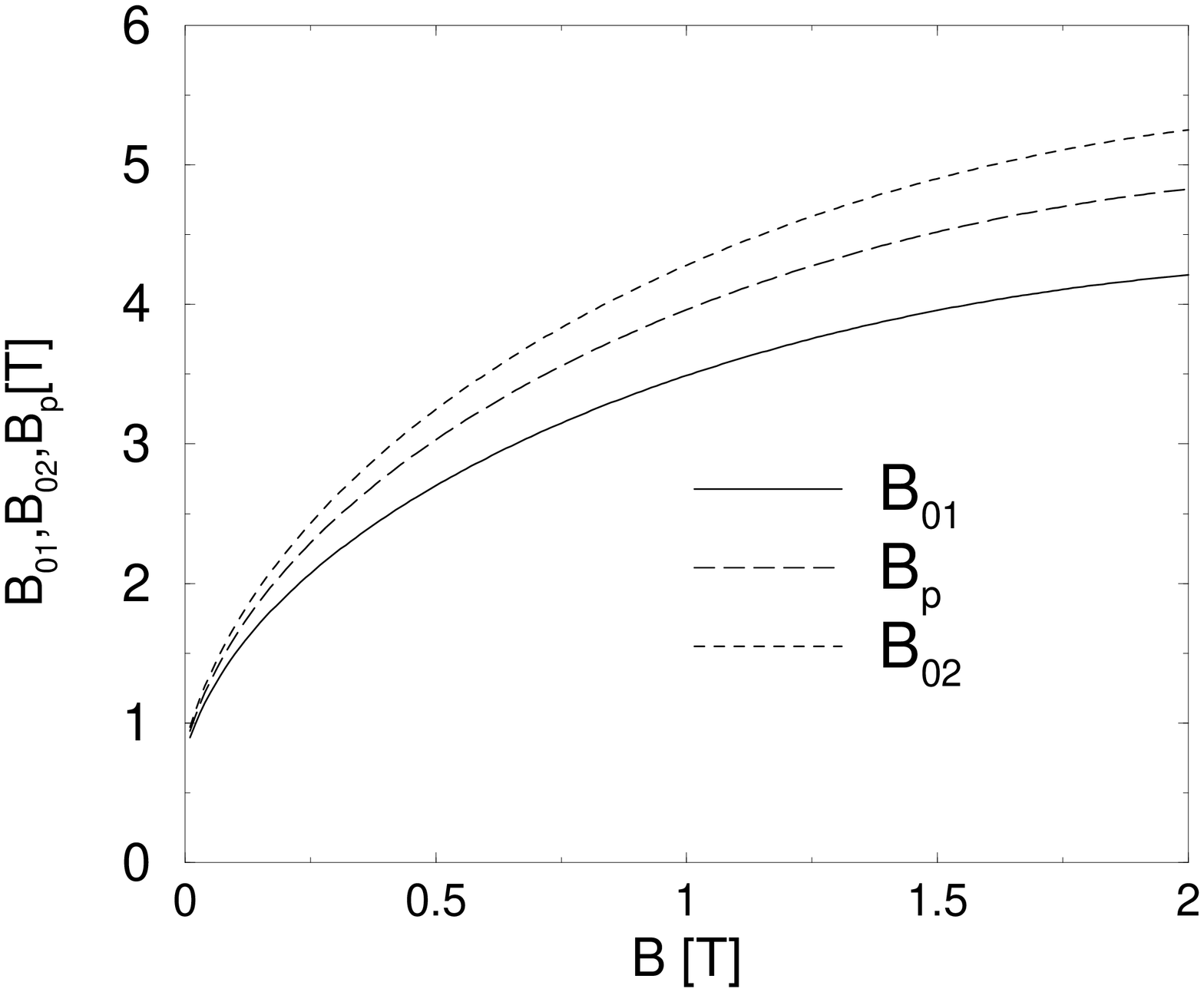,width=0.3\textwidth,clip=,angle=-90}
\end{center}
\caption{Field scales $B_{0m}$ and $B_p$ for the variation of the 
JPR resonances $\omega_m$ and the pole $\omega_{\rm pole}$ in the limit 
of high fields $B \gg B_{\lambda}$, cf. Eq.~(\ref{caseII}) 
($B_{\rm melt} = 2$ T, $\lambda_{ab}=1700$ \AA, $T=2.3$ K).
\label{BtypeII}
}
\end{figure}

In Fig.~\ref{Bcomp} it is seen that in the low field limit 
$B \ll B_{Jm}$ as determined in 
Eq.~(\ref{caseI}) the field scale 
$B_p$ determining the variation of the peak frequency 
$\omega_{\rm pole}$ in ${\rm Im}(\omega \epsilon_{\rm eff})$ 
is field independent. 
In contrast to this, the self-consistent solution of Eq.~(\ref{caseII})
valid for $B\gg B_{\lambda}$
introduces a variation of $B_p$ with $B$.
In Fig.~\ref{BtypeII} the  field scales $B_{0m}$ and $B_p$ responsible for 
the variation of the JPR peaks $\omega_m$ in $L(\omega)$ and of the pole 
$\omega_{\rm pole}$ in ${\rm Im}(\omega \epsilon_{\rm eff})$ are compared. 
The sensitivity of the magnetic field scale $B_{0m}$ to the choice of the
correct melting field is low.

\begin{figure}
\begin{center} 
\epsfig{file=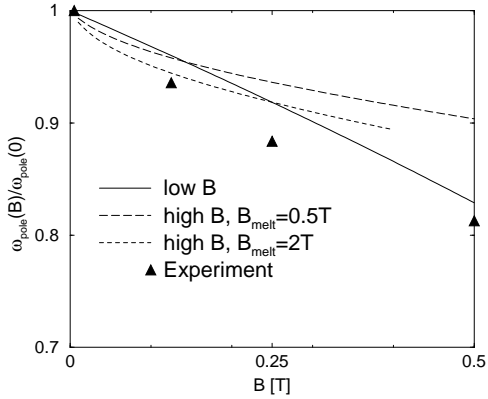,width=0.3\textwidth,clip=,angle=-90}
\end{center}
\caption{
Frequency $\omega_{\rm pole} / \omega_{\rm pole} (0) $ of the 
peak in ${\rm Im} (\omega \epsilon_{\rm eff})$ in the low and high field 
limit $B \ll B_{Jm}, B_{\lambda}$ or $B \gg B_{Jm}, B_{\lambda}$ 
using Eq.~(\ref{caseI}) or Eq.~(\ref{caseII}) respectively 
($T=2.3$ K). Best fit of the experimental data in Ref. [6] is generally 
obtained for the penetration depth $\lambda_{ab}=1850$ \AA, except for the
case, when a too large melting field $B_{\rm melt}=2$ T is chosen for comparison, which 
requires $\lambda_{ab} = 2230$ \AA. 
\label{omega23K} 
}
\end{figure}

\begin{figure}
\begin{center} 
\epsfig{file=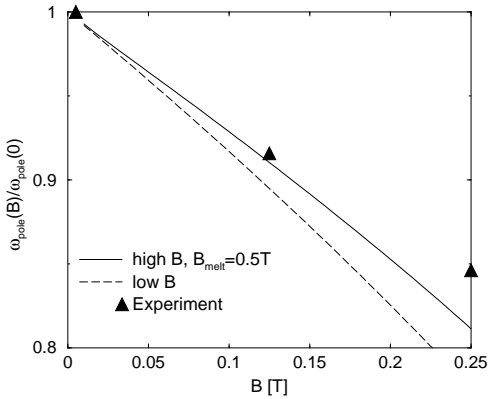,width=0.3\textwidth,clip=,angle=-90}
\end{center}
\caption{Best fit of $\omega_{\rm pole} / \omega_{\rm pole} (0)$ for high 
temperature $T=10$ K using the 
low and high field limits in Eq.~(\ref{caseI}) ($\lambda_{ab}=1600$ \AA) 
and Eq.~(\ref{caseII}) ($\lambda_{ab}=1100$ \AA, $B_{\rm melt}= 0.5$ T).
\label{omega10K}
}
\end{figure}

We now compare the calculated frequencies $\omega_m(B)$ 
with the experimental results in
Ref.~\onlinecite{pimenov}, 
where the eigenfrequencies $\omega_{0m} (B)$ are extracted from the
zeros of 
the effective dielectric function $\epsilon_{\rm eff}$ and the
characteristic
frequency $\omega_{\rm pole}$ from the peak in ${\rm Im} (\omega \epsilon_{\rm
eff})$.

Firstly, we use the normalized values of 
$\omega_{\rm pole}(B) /\omega_{\rm pole}(0)$, as they
do not depend on the choice of the charge coupling parameter $\alpha$, see
Figs.~\ref{omega23K} and \ref{omega10K}. For low fields the results derived in the cases 
$B \ll B_{\lambda}, B_{Jm}$ (Eq.~(\ref{caseI})) and $B \gg  B_{\lambda}, B_{Jm}$ 
(Eq.~(\ref{caseII})) are close, but deviate at 
fields approaching the melting field $B_{\rm melt} \approx 0.5$ T,
which is chosen in accordance with the observed phase transition to
the vortex liquid. The agreement with the experimental data is
worsening near the vortex solid-liquid transition.

For low temperatures, cf. Fig.~\ref{omega23K}, agreement with the experimental
data can be obtained for $\lambda_{ab}=1850$ \AA, while at high $T=10$ K we get 
$\lambda_{ab}=1600$ \AA \ \ using Eq.~(\ref{caseI}) or
$\lambda_{ab}= 1100$ \AA \ \ ($B_m=0.5$ T) from  Eq.~(\ref{caseII}). 
Following from estimates for $B_{\lambda}$ 
the latter case seems to be more adequate.  
For higher temperatures, cf. Fig.~\ref{omega10K},  the assumptions of our
theory are expected to be better justified and in this case 
the obtained value for $\lambda_{ab}=1100$ \AA \ \
is comparable to the ones reported for other layered superconductors.  

 \begin{figure}
\begin{center} 
\epsfig{file=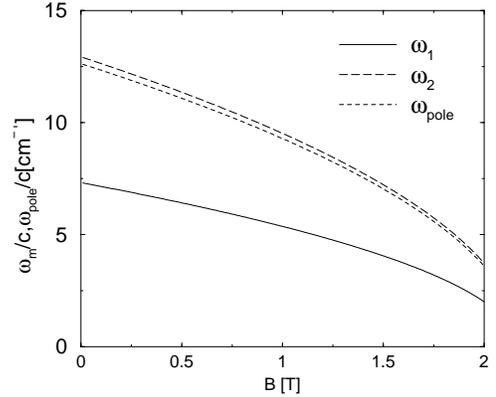,width=0.3\textwidth,clip=,angle=-90}
\end{center}
\caption{JPR peaks $\omega_{m}(B)$ in $L(\omega)$ and 
$\omega_{\rm pole}(B)$ in ${\rm Im} (\omega \epsilon_{\rm eff})$ 
using the low field limit ($\lambda_{ab}=1700$ \AA, $T=2.3$ K, $\alpha=0.4$, 
$\omega_{c1} (B=0) /c = 6.6$ cm$^{-1}$, 
$\omega_{c2} (B=0) /c = 8.9$ cm$^{-1}$, cf. Ref. [6,15]). 
\label{omegalongI}
}
\end{figure}

\begin{figure}
\begin{center} 
\epsfig{file=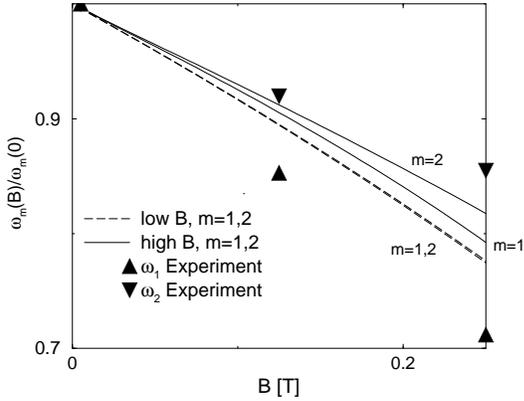,width=0.3\textwidth,clip=,angle=-90}
\end{center}
\caption{Best fit of 
$\omega_{0m}(B) / \omega_{0m}(0) $  from Ref. [6]
for high temperatures $T=10{\rm K}$ using the low (dashed, $\lambda_{ab}=1600$\AA)
 and high (solid, $\lambda_{ab}=1100$\AA, $B_{\rm melt}=0.5{\rm T}$) field approximations,
Eqs.~(\ref{caseI}) and (\ref{caseII}) respectively ($\alpha=0.4$).
\label{omegalongIIlowT}
}
\end{figure}

Assuming the charge coupling $\alpha \approx 0.4$ as estimated in 
Ref.~\onlinecite{ourprb} from the vortex liquid state and the shape of the
loss function for $B=0$, the resonance frequencies $\omega_m (B,\alpha)$ in $L(\omega)$
and $\omega_{\rm pole}$ are shown in Fig.~\ref{omegalongI}. 
The experimental data for the normalized JPR resonances 
$\omega_{m} (B) / \omega_{m}(0) $ for high temperatures can be fitted with the
same choice of parameters as for $\omega_{\rm pole}$, cf.  Fig.~\ref{omegalongIIlowT}. 

To conclude, thermal fluctuations in the vortex solid state describe 
satisfactorily the dependence of the plasma frequencies in the 
Josephson coupled system SmLa$_{1-x}$Sr$_x$CuO$_{4-\delta}$ with two 
different alternating junctions on a perpendicular magnetic field.  
From the comparison of our theoretical results and the experimental data we 
estimate the in-plane London penetration length $\lambda_{ab}\approx
1100 $ \AA \ \ at high temperatures $T=10$ K, 
which is similar to other layered superconductors. 

The authors thank Dirk van der Marel and A. Pimenov for useful 
discussions and providing the experimental data.   
This work was supported by the Los Alamos National
Laboratory under the auspices of the US DOE and by the Swiss National
Science Foundation through the National Center of Competence
in Research "Materials with Novel Electronic Properties-MaNEP".

\end{multicols}

\end{document}